\documentstyle[aas2pp4,psfig2]{article}

\slugcomment{Submitted to ApJ Letters}

\lefthead{Hopkins et al.}
\righthead{Microjansky sources at 1.4\,GHz}

\begin{document}

\title{Microjansky sources at 1.4\,GHz}

\author{A. Hopkins\altaffilmark{1}} 

\affil{Australia Telescope National Facility, PO Box 76 Epping, NSW 1710,
  Australia.}

\and

\author{J. Afonso}

\affil{Astrophysics Group, Imperial College, Blackett Laboratory, Prince
  Consort Rd, London SW7 2BZ, UK}

\and

\author{L. Cram}

\affil{School of Physics, University of Sydney, NSW 2006, Australia}

\and

\author{B. Mobasher}
\affil{Astrophysics Group, Imperial College, Blackett Laboratory, Prince
Consort Rd, London SW7 2BZ, UK}

\altaffiltext{1}{School of Physics, University of Sydney,
NSW 2006, Australia.} 

\begin{abstract}
  We present a deep 1.4\,GHz survey made with the Australia Telescope
  Compact Array (ATCA), having a background RMS of $9\,\mu$Jy near the
  image phase centre, up to $25\,\mu$Jy at the edge of a $50'$ field of
  view. Over 770 radio sources brighter than $45\,\mu$Jy have been
  catalogued in the field. The differential source counts in the deep field
  provide tentative support for the growing evidence that the microjansky
  radio population exhibits significantly higher clustering than found at
  higher flux density cutoffs. The optical identification rate on CCD
  images is approximately 50\% to $R=22.5$, and the optical counterparts of
  the faintest radio sources appear to be mainly single galaxies close to
  this optical magnitude limit.
\end{abstract}


\section{Introduction}

Our understanding of the faintest radio sources has advanced over the last
decade through increasingly sensitive radio surveys and through follow-up
work at optical and infrared wavelengths. The appearance of a `new'
population of radio galaxies at millijansky and sub-millijansky levels,
first revealed by the changing slope of the 1.4\,GHz radio source count
distribution, has been confirmed by spectroscopic and multicolour studies
(\cite{Wind:85,Thu:87,Ben:93,Wind:94,Ham:95,Hop:98,Rich:98}). The
population comprises star forming galaxies whose numbers increasingly
dominate classical (AGN-powered) radio sources as the flux density limit
falls below 1\,mJy. The evolution of these galaxies, and the role of
interactions and mergers in the population, are only partially understood.

The number of catalogued microjansky radio sources at 5 and 8\,GHz is a few
hundred, and at 1.4\,GHz there are few catalogued sources fainter than
$100\,\mu$Jy. The {\em Phoenix Deep Survey} (PDS) aims to increase the
number of known 1.4\,GHz sources fainter than $100\,\mu$Jy. The PDS covers
a 2\arcdeg\ diameter field selected to lie in a region of low optical
obscuration and devoid of bright radio sources (\cite{Hop:98}). The PDS
provides a large, homogeneous sample of 1.4\,GHz sources which, through
multiwavelength observations, is being used to investigate starburst and
post-starburst galaxies in the the faint radio population
(\cite{Hop:97,Hop:98,Cram:98,Hop:99,Age:99,Mob:99}). We present here the
source counts and optical identifications in a Phoenix mosaic image which
contains many sources fainter than $100\,\mu$Jy.
 
\section{Observations, data reduction and imaging}
 
Earlier radio observations of the PDS at 1.4\,GHz provided images with a
$5\sigma$ sensitivity of $300\,\mu$Jy over the 2\arcdeg\ Phoenix Deep Field
(PDF), and $100\,\mu$Jy in a sub-region of $36'$ diameter, referred to as
the Phoenix Deep Field Sub-region (PDFS) (\cite{Hop:98}). Additional
observations were made in 1997 November and December, using the ATCA-6C
configuration in mosaic mode with 32 channel, 128\,MHz bands centred at
1.380\,GHz and 1.472\,GHz. The mosaic has seven pointings, six arranged
hexagonally around a pointing centred on the PDFS
[$\alpha(J2000)=01^{\rm{h}}~11^{\rm{m}}~13\fs0$,
$\delta(J2000)=-45^{\circ}~45'~0\farcs0$]. A integration of 122 hours was
spent on the mosaic. The primary calibrator was B1934-638. The phase
calibrator, B0153-410, was observed for ten minutes every hour. An
additional 42 hours of data obtained in 1994 September on the central PDFS
pointing were included in the imaging.
 
The {\sc miriad} (Multichannel Image Reconstruction, Image Analysis and
Display) data analysis package was used to edit and calibrate the data, as
with previous PDS observations (\cite{Hop:98}). An artifact at the phase
centre of earlier PDS images due to self-interference from the sampler
clocks is no longer present.
 
An imaging cell of $2 \times 2$ arcsec$^2$ per pixel was used, providing at
least 3 pixels across the shortest FWHM of the synthesised beam. Naturally
and uniformly weighted images were produced. The RMS noise in the naturally
weighted image is lower, but the synthesised beam larger ($\alpha \times
\delta \approx 6'' \times 12''$ compared with $5'' \times 7''$ for uniform
weighting). Sources have been detected in the naturally weighted image,
and individual sources (when detected) have been investigated using the
uniformly weighted image.

\begin{figure}
\centerline{\psfig{figure=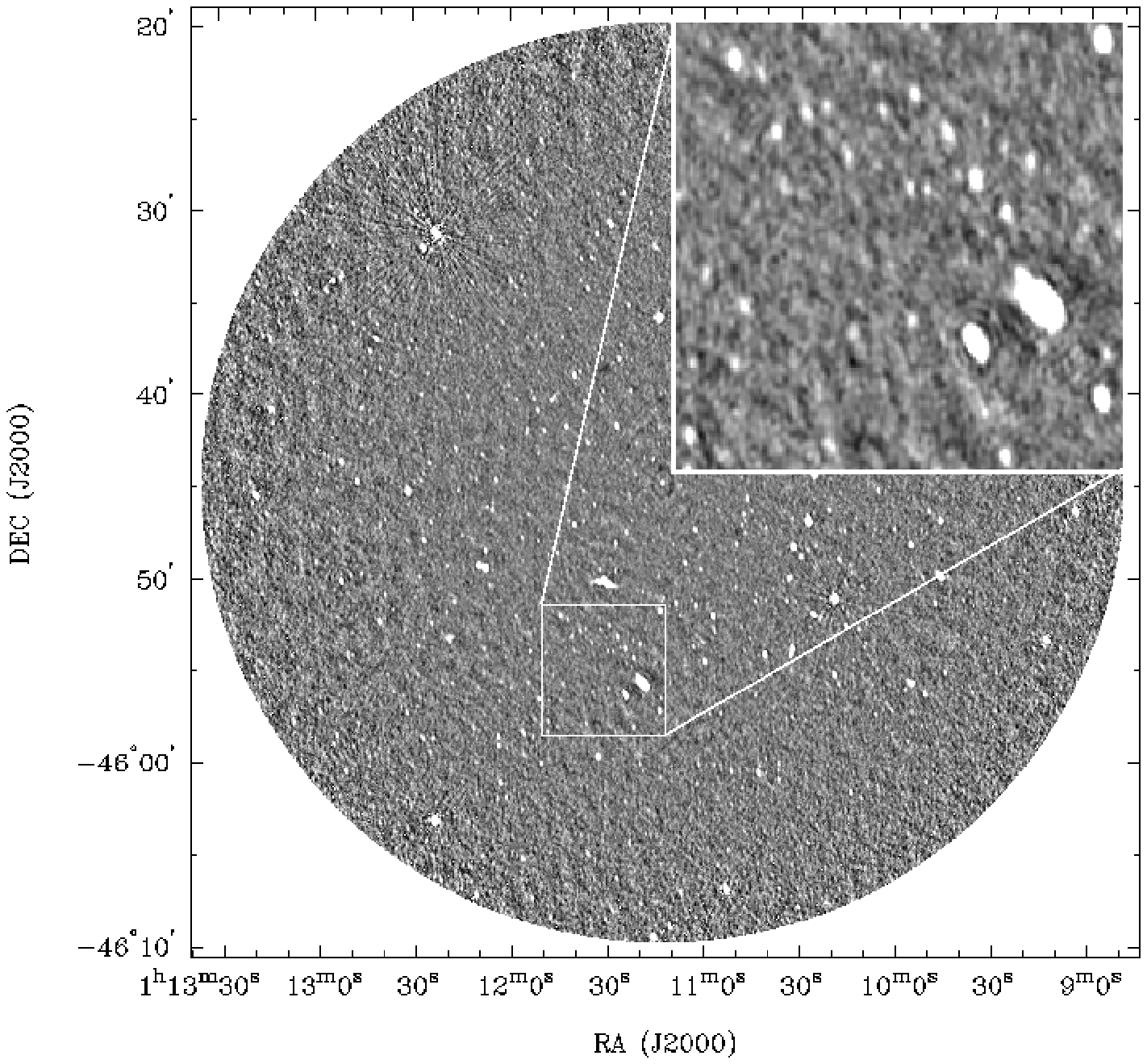,height=7cm,rotate=0}}
\caption{Grey scale image of the central $50'$ diameter region of the mosaic.
  The brightest source (in the NE of the image) is 23\,mJy, the faintest
  $45\,\mu$Jy. The insert shows greater detail for a small portion of the
  image.
  \label{deepimage}}
\end{figure}

Mosaic components at each pointing were made 2100 pixels ($70'$)
square, to allow correct {\sc clean}ing of the sidelobes of out-of-field
sources. Although the PDF and PDFS fields were chosen to avoid bright
sources, the two westernmost pointings were affected by the sidelobes of a
strong source outside the field. These were removed by imaging, modelling
and removing this source (using the {\sc miriad} task {\sc uvmod}) when
constructing the images for those pointings. The images for each pointing
were {\sc clean}ed individually, in each case reaching the thermal noise,
before being combined to form the final mosaic. The central 50$'$ diameter
of the mosaic is shown in Figure~\ref{deepimage}.

Because the final image is a mosaic in which the individual pointings have
different integration times, spatial overlap, and primary beam attenuation,
the distribution of noise in the image is not uniform. The RMS noise is a
minimum at the centre ($9\,\mu$Jy) rising uniformly to $25\,\mu$Jy at a
radius of $25'$. The RMS noise at the centre is a factor of 2.5 lower than
the previous PDFS image (\cite{Hop:98}), consistent with the 4-fold
increase in integration time and the absence of the phase-centre artifact
present in the earlier observations. The dynamic range of the image can be
described in two ways: the brightest source (in the NE, 22.7 mJy) lies more
than a factor of 100 above the RMS noise of its surrounds, while the ratio
of the flux densities of the brightest and faintest reliably detected
sources exceeds 500:1.

\section{Radio and optical sources}

\subsection{1.4\,GHz sources}

The mosaic achieves wide-field coverage and good sensitivity at the price
of having an unavoidably non-uniform noise distribution. The detection and
statistical characterisation of sources is thus a more complex procedure
than it is in single-pointing interferometer images. An important
auxiliary in this process is the noise image, in which each pixel is
assigned a value equal to the theoretical $1\sigma$ noise level accounting
for the observing time, mosaic overlap and primary beam attenuation. The
{\sc miriad} task {\sc sfind} (\cite{Hop:98}) was used to compile a
list of sources which have a peak flux density greater than 4 times the
value in the noise image at the same position, lying in the central $50'$
diameter disk of the mosaic. This disk is the area bounded by the
$25\,\mu$Jy ($1\sigma$) contour. Over 770 sources within this region having
a peak flux density brighter than the {\em local} value of $4\sigma$
survive visual inspection in the interactive confirmation phase of {\sc sfind}.
There is some deviation from Gaussian statistics in the distribution of
negative pixel values ($\lesssim 0.1$\% of pixels lie in a non-Gaussian
tail associated with visible sidelobes and the edges of the image) which
evidently exists in the positive pixel values as well, albeit masked by
true sources. However, the procedure we have adopted provides a very robust
estimate of the noise level above which we claim detections, and excludes
essentially all spurious positive candidates.

\begin{figure}
\centerline{\psfig{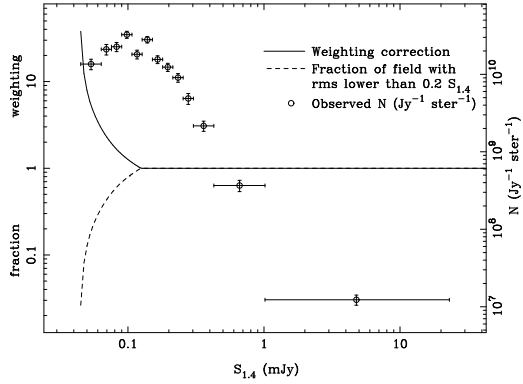}}
\caption{Points (right scale) show the {\em raw} source counts, with
  vertical error bars denoting Poisson variation and horizontal bars
  denoting the range of source flux density averaged together. The fraction
  of the mosaic area covered by noise with $5\sigma > S_{1.4}$ is shown as
  a dashed curve, and the weighting correction derived from this is shown a
  a solid curve. Approximately 10\% of the image has $5\sigma < 50\,\mu$Jy.
  \label{rawcounts}}
\end{figure}

\begin{figure}
\centerline{\psfig{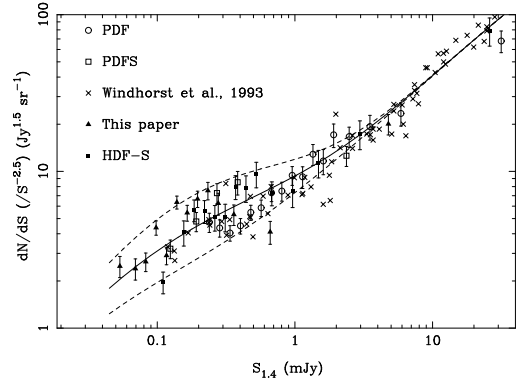}}
\caption{Normalised differential 1.4\,GHz source counts. Crosses
  are a compilation of data from other surveys (\protect\cite{Wind:93}).
  Open circles, open squares and filled triangles are from radio
  observations of the PDF. Filled squares are from the Hubble Deep Field
  South region (\protect\cite{Nor:99}).
  \label{scounts}}
\end{figure}

The construction of the source count distribution takes account of two
significant corrections (c.f. \cite{Hop:98}). First, the use of a peak
detection algorithm to locate sources will underestimate the total number
of sources to a given total flux density. A correction based on the
statistical distribution of source sizes has been applied. Secondly, it is
necessary to correct the raw counts at any noise level for the fraction of
the total area over which such sources could be detected above $4\sigma$.
Figure~\ref{rawcounts} illustrates the magnitude of this latter correction,
while Figure~\ref{scounts} shows the corrected counts.

Questions of sensitivity, completeness, spurious sources, and source
confusion are important in a survey exploited close to its limit, but are
somewhat difficult to answer in view of the non-uniform noise properties of
the mosaiced images and the fact that some but not all of the sources are
resolved. The observed field contains approximately $10^5$ independent beam
areas and some 750 sources, resulting in over 100 beam areas per source except
in the few small regions of highest source density within the highest
sensitivity area (only a few percent of the field). Hence we have
not quite reached the sensitivity and source density at which confusion would
become a serious problem. We have established that there are
only a few deviations from a Gaussian distribution in the negative pixels,
due to fluctuations similar to those edited out in {\sc sfind}. We thus
expect approximately 10 spurious sources due to noise statistics with our
$4\sigma$ peak flux cut-off. The catalogue completeness can be described
in terms of the factors required to correct for the primary beam weighting
and the presence of resolved sources (\cite{Hop:98}). The survey has
50\% completeness at that peak flux density which requires a factor of 2
correction for these effects. The peak flux density of the 50\%
completeness limit varies with radial position, $r$. It is $54\,\mu$Jy for
$r < 12'$; $69\,\mu$Jy for $r < 21'$; and $83\,\mu$Jy for $r < 26'$.
The faintest catalogued source has an integrated flux density of $45\,\mu$Jy.

\subsection{Optical counterparts}

The optical catalogue is derived from Anglo-Australian Telescope (AAT)
prime-focus CCD observations made in the Johnson-Kron-Cousins $R$-band
(\cite{Age:99}). The most probable optical candidate (where one exists) is
chosen by searching a radius of $5''$, and selecting the source with the
least probability of being an accidental alignment (given the known
surface density of sources as bright or brighter than the candidate)
provided that that probability $<5$\%. While some true associations may
be missed by this method it minimises spurious associations. Of the 773
detected radio sources, 52\% have been optically identified. As
explained by \cite{Age:99} the 50\% completeness limit of the
optical survey is $R=22.5$ although sources fainter than this have been
reliably detected and are included in the analysis.

\begin{figure}
\centerline{\psfig{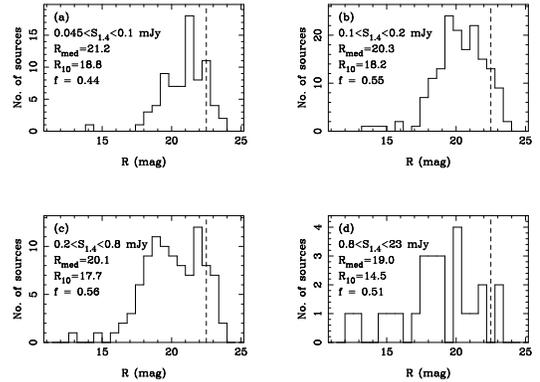}}
\caption{Histograms of optical counterpart magnitudes for
  selected ranges of 1.4\,GHz flux density. The median $R$-band magnitude
  of the optical counterparts in the given flux density range is shown for
  each panel. The fraction $f$ of optical counterparts identified for the
  flux density range is also shown. The vertical dashed line indicates the
  completeness limit of the sample at $R=22.5$.
  \label{opmags}}
\end{figure}

Figure~\ref{opmags} presents histograms of the distribution of apparent
$R$-band magnitude as a function of the radio source flux density. The median
optical magnitude ($R_{\rm med}$) of the detected optical counterparts
falls with decreasing radio flux density. The magnitude bounding the first
decile ($R_{10}$) also decreases to fainter optical magnitudes as the radio
brightness declines. The fraction of sources in each flux density bin with
identified optical counterparts, $f$, is smaller for the faintest bin, to
our fixed limiting optical magnitude.

\section{Discussion}

Over 50\% of the $S_{1.4} > 100\,\mu$Jy radio sample is optically
identified to $R=22.5$, with a lower rate (44\%) for fainter radio sources.
Visual inspection shows that the faint ($<100\,\mu$Jy) sources are mostly
identified with single optical galaxies, rather than interacting systems.
This may reflect the fact that these identifications are made quite close
to the limit of the CCD images, or it may indicate that the very faint
radio counterparts to a fixed optical magnitude are located in galaxies
different from harbouring slightly brighter sub-mJy radio sources (c.f.
\cite{Grup:97}).

Figure~\ref{scounts} reveals considerable scatter in different
determinations of the 1.4\,GHz source counts below about 1 mJy. Counts
from different surveys fluctuate by amounts large compared with the Poisson
errors and with the respective image variances. The effect can be seen
between counts in the $50'$ diameter survey area presented here and those in
the entire 2\arcdeg\ Phoenix field. It seems unlikely that limited
dynamic range, sidelobe confusion, or the blending of faint sources to
mimic single brighter sources could account for the effect within the
Phoenix data set or within the other catalogues.

It is known that large samples of radio sources brighter than $\approx
10\,$mJy at 1.4\,GHz show no or weak evidence of clustering (e.g.
\cite{Bal:98}). However, fainter samples may have revealed clustering
(\cite{Cres:96}), and it is predicted (\cite{BW:95}) that a $30'$ diameter
survey with $S_{\rm min}<100\,\mu$Jy might well reveal fluctuations above
Poisson statistics. Indeed, Benn \& Wall caution that deep radio surveys
with pencil beams may not be representative. Since the clustering scale
responsible for the observed fluctuations is likely to be
$\rho\sim(100\,h^{-1}\,{\rm Mpc})^{3}$, extending the Phoenix survey at its
current sensitivity limit to a field of several degrees could provide
constraints to large-scale structure on scales intermediate between those
probed by microwave background experiments and by optical redshift surveys.

\section{Conclusions}

A new 1.4\,GHz radio survey with the ATCA has been completed as part of the
{\em Phoenix Deep Survey}. A catalogue of 773 sources with
$S_{1.4}\gtrsim45\,\mu$Jy has been compiled. This sample is homogeneously
selected and thus uniform across the entire flux density range. The new
sample provides an opportunity to investigate a large number of faint radio
galaxies through multicolour photometry and spectroscopy.

Statistical fluctuations in observed 1.4\,GHz source counts below
$500\,\mu$Jy provide a tantalising hint that very sensitive large area
1.4\,GHz surveys may be used to provide constraints to cosmic structure on
the scale $\rho\sim(100\,h^{-1}\,{\rm Mpc})^3$, intermediate between scales
probed by microwave background experiments and by optical redshift surveys.

\acknowledgments

We thank the ATCA HDF-S team for generously making their 1.4\,GHz source
count data available. We are grateful to Eric Richards for helpful
comments and discussion. JMA gratefully acknowledges support in the form
of a scholarship from Funda\c{c}\~ao para a Ci\^encia e a Tecnologia
through Programa Praxis XXI. The Australia Telescope is funded by the
Commonwealth of Australia for operation as a National Facility managed by
CSIRO.


\clearpage
 
\end{document}